# People's Views on Smiley Emoji on WeChat Platform, Positive Emotion and Sarcasm

Yuyang Shi[1]

[1] Beijing University of Posts and Telecommunications

## Abstract

Emoji can easily lead to misunderstanding. On WeChat, [Smile] is often used by the elderly to express positive emotions such as joy and praise, but young people often regard [Smile] as sarcasm. In this case, misunderstandings can easily occur. In order to confirm the existence of this phenomenon and analyze people's understanding of emoji, this study selected 20 smiley emoji on WeChat platform and directly asked people's opinions on these emoji through questionnaires, including whether they recommend others to use, and rating of emotion joy, praise, sarcasm. Analyses were performed based on the total sample and by age group. The results showed that [Smile] and [Bye] were interpreted in opposite ways by older people and younger people. Older people thought they represented positive emotion (joy, praise), while younger people thought they represented sarcasm. Such emojis are easy to misunderstand. On the other hand, some emojis like [Yeah!], both the old and the young have a similar understanding that it represents a positive emotion.

# Introduction

## Motivation

On the WeChat platform, some emojis are easy to cause misunderstanding, such as [Smile] and [Bye], these two emojis are often used on the WeChat platform to indicate sarcasm, angry and to end a conversation unfriendly (Xu, 2023). However, this is not the case for everyone. Many old people or people who are not familiar with emoji-related culture still use emojis, especially [Smile] emoji, because they do not understand the negative connotation, which will cause misunderstanding (Lu & Wu, 2022). [Smile] is often interpreted as a sign of sarcasm. For example, a teacher who sends his student "You're doing a great job [Smile]", a message that at first glance seems sarcastic to the student. The student can further understand the real meaning of the teacher in the follow-up communication, but this will cause misunderstanding and undoubtedly increase the communication cost. If [Smile] is used in peer communication, and peers often believe that they share the same understanding, a young person who does not understand the negative connotation of [Smile] will send [Smile] to his peer, the recipient will be more convinced that it is an unfriendly sarcasm message, which will lead to greater misunderstanding, which will damage their relationship.

Among young people, instant messages often need to be attached with an emoji, and many people think that messages without emoji are offensive, at the same time, emoji are not systematic and vague, and misunderstandings are easy to occur (Lu & Wu, 2022). This will make people more cautious about using emoji. People need to carefully choose emojis that are not easy to cause misunderstanding, which brings difficulties to people, especially those who are not familiar with emoji. Choosing the right emoji for expression can make people struggle. Research is needed to investigate people's emoji usage habits in instant communication, help people use emoji better and more smoothly, and reduce the misunderstanding caused by emoji.

People with visual impairments may also need this research. In the voicemoji study, researchers investigated the most needed emoji input needs of visually impaired people and developed a voice input system to help them input emoji and the voice input system recommends emoji through the Dango API and DeepMoji model (Zhang et al., 2021). The training data for these recommendation models is generally public content, such as tweets, which may not reflect people's behavior in private instant communication. It is necessary to study the use of emoji in private instant messaging tools, which can provide valuable data and more appropriate data labels for model training.

**Related Work**

Certain facial expressions usually express certain feelings, which is also true across cultures, and New Guineans who have little exposure to Western culture can correctly identify the emotions expressed by Western faces (Ekman & Friesen, 1971). In subsequent studies on Cameroonians, Tanzanians and Japanese, the Japanese could correctly identify the emotions expressed by emoji and real human faces, while the Cameroonians and Tanzanians could identify the emotions expressed by real human faces but could not understand the emotions represented by emoji (Takahashi et al., 2017). It can be seen that in the cross-cultural situation, the research results on emotion recognition of faces are the same. It can be assumed that under the same cultural background, or the same network cultural background, people have the same understanding of emoji.

Researchers have been looking for the correspondence between emojis and emotions. The emotions represented by emojis can be analyzed by algorithms. In one study, researchers classified emojis into joy, sad, and angry using Naive Bayes Algorithm (Sendari et al., 2020). The display of emojis in Android and iOS systems is different. In a study, researchers found emojis that express similar emotions in Android and iOS systems, and reply with emojis that express similar emotions will have better communication effects (Tigwell & Flatla, 2016). Some statistical studies on emoji usage can reveal people's emoji usage habits. A study based on data from Twitter showed that "Face with tears of joy" was the most used emoji, and several other highly used emojis mimicked facial expressions (Ljubešić & Fišer, 2016). A study based on Weibo data found that most of the emojis in the text are at the end of the text, and most of the text uses only one emoji (Yang & Liu, 2021). However, these studies have certain limitations. The data sources of these studies are all public content, and people's emoji usage habits in public posts and private instant messages may be different, so research on people's emoji usage habits in instant messages is necessary.

Some research has focused on sarcasm and emoji. Emoji can effectively express sarcasm, and in ASCII emoji sarcasm is often expressed using wink and tongue (Thompson & Filik, 2016). Eye-tracking experiments, video experiments, and story experiments show that compared with young people, sarcasm is more difficult for the elderly to understand (Howman & Filik, 2020; Phillips et al., 2015). A study of wink emoji found that emoji can help people understand sarcasm, and sentences with emoji are equally well understood by older people as by younger people (Garcia et al., 2022).

In a study of emoji on Android, Apple, and Samsung platforms, participants rated each emoji according to the emotion label given by the researcher to assess users' understanding of the emotion represented by the emoji (Franco & Fugate, 2020). This study will refer to their method and ask the participants to give their opinions on emojis directly.

Table 1

23 WeChat emojis in experiment Chinese and English label comparison table

| No. | 1 | 2 | 3 | 4 | 5 | 6 | 7 | 8 | 9 | 10 |
|---|---|---|---|---|---|---|---|---|---|---|
| Visual presentation | 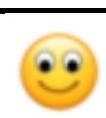 | 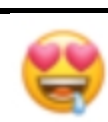 | 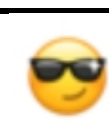 | 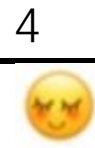 | 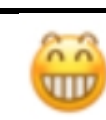 | 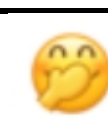 | 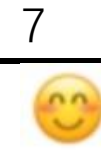 | 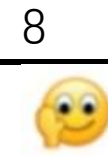 | 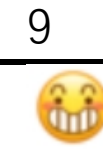 | 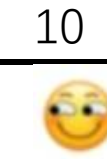 |
| Chinese | 微笑 | 色 | 得意 | 害羞 | 呲牙 | 偷笑 | 愉快 | 再见 | 坏笑 | 奸笑 |
| English | Smile | Drool | CoolGuy | Shy | Grin | Chuckle | Joyful | Bye | Trick | Smirk |

| No. | 11 | 12 | 13 | 14 | 15 | 16 | 17 | 18 | 19 | 20 |
|---|---|---|---|---|---|---|---|---|---|---|
| Visual presentation | 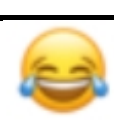 | 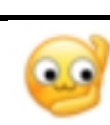 | 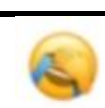 | 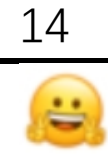 | 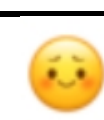 | 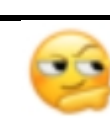 | 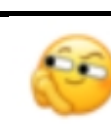 | 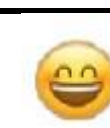 | 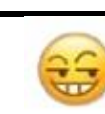 | 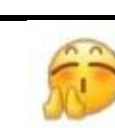 |
| Chinese | 破涕为笑 | 嘿哈 | 捂脸 | 耶 | 皱眉 | 机智 | 社会社会 | 笑脸 | 阴险 | 鼓掌 |
| English | Lol | Hey | Facepalm | Yeah! | Concerned | Smart | Respect | Happy | Sly | Clap |

Table 1 (continued)

| No. | 21 | 22 | 23 |
|---|---|---|---|
| Visual presentation | 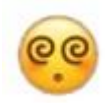 | 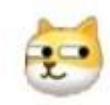 | 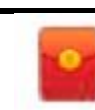 |
| Chinese | 晕 | 旺柴 | 红包 |
| English | Dizzy | Doge | Packet |

**Pre-experiment Survey**

There are 23 emojis in the emoji set of the experiment. Emojis 1-20 are experimental emojis, and emojis 21-23 are fillers.

For smiliness of expressions 1-20 an experiment with 10 participants was conducted to evaluate smiliness of expressions by a question.

*Use 1-5 to rate smiliness of the following emojis.*

*Very not smiley 1 2 3 4 5 Very smiley*

Table 2
Smiliness rating statistics of emojis from #1-20

| No. | 1 | 2 | 3 | 4 | 5 | 6 | 7 | 8 | 9 | 10 |
|---|---|---|---|---|---|---|---|---|---|---|
| Visual presentation | 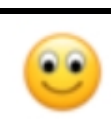 | 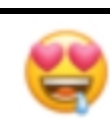 | 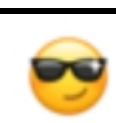 | 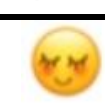 | 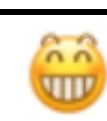 | 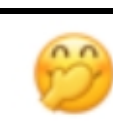 | 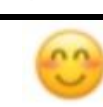 | 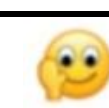 | 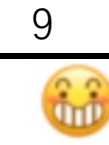 | 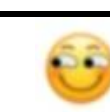 |
| Avg | 4.20 | 3.60 | 3.40 | 3.50 | 4.70 | 4.40 | 4.30 | 3.70 | 4.40 | 4.30 |
| SD | 1.317 | .966 | .966 | .850 | .675 | .516 | 1.059 | 1.418 | .699 | .675 |
| Kurtosis | 3.607 | -.623 | -.623 | .107 | 4.765 | -2.277 | 1.258 | -.378 | -.146 | -.283 |
| skewness | -1.913 | .111 | -.111 | .000 | -2.277 | .484 | -1.444 | -.801 | -.708 | -.434 |

| No. | 11 | 12 | 13 | 14 | 15 | 16 | 17 | 18 | 19 | 20 |
|---|---|---|---|---|---|---|---|---|---|---|
| Visual presentation | 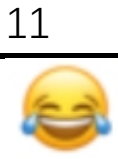 | 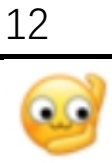 | 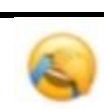 | 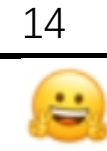 | 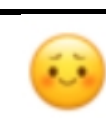 | 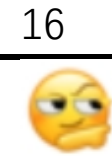 | 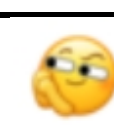 | 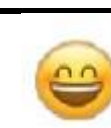 | 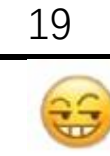 | 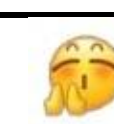 |
| Avg | 4.10 | 2.40 | 3.00 | 4.50 | 3.00 | 2.50 | 2.30 | 4.70 | 3.30 | 2.20 |
| SD | 1.101 | .699 | 1.491 | .527 | 1.155 | 1.080 | 1.059 | .483 | 1.059 | 1.229 |
| Kurtosis | -.522 | -.146 | -1.334 | -2.571 | .080 | -1.032 | -.406 | -1.224 | -1.238 | -1.461 |
| skewness | -.863 | -.780 | .000 | .000 | .000 | .000 | .659 | -1.035 | -.042 | .431 |

#5[Grind] gets a higher score of smiliness. Its kurtosis is greater than zero and has a large value and shows negative skewness.

Smiliness scores were lower for emoji #12, 16, 17, and 20.

# Method

## Participant

The experiment was conducted through the Questionnaire Star platform. Participants were able to open the questionnaire by following the link to answer the questions. The questionnaire uses personal information questions at the beginning to filter the appropriate data in the data analysis.

A total of 32 answers were received, and after screening, 29 answers whose Internet culture background was Chinese culture were used as samples for analysis.

## Task

The questionnaire was divided into a personal information part and two experimental parts. In the personal information section, participants were asked to select their gender, age, education, occupation, Internet culture background, and knowledge of emoji.

The emoji set used in the experiment is the 23 emoji in Table 1. Emoji #1-20 are experimental emoji, and emoji #21-23 are fillers. Three filler emojis were placed at the beginning, middle and end of each experimental part.

In the first part of the experiment, participants were presented with an emoji in each question and asked to choose one of three options to say whether they would recommend the emoji or not.

*Whether users unfamiliar with the emoji are advised not to use it?*

*<emoji>*

a) *It can be used with confidence in most cases, without causing misunderstanding.*
b) *Do not use the emoji unless the user is very sure of his or her intention in using it.*
c) *Not sure about this one.*

In the second part of the experiment, participants were asked to rate an emoji from three perspectives: joy, praise, and sarcasm.

*How much do you agree with this emoji to express joy, praise, and sarcasm?*

*<emoji>*

- *Joy Strongly disagree1-5 Strongly agree*

- *Praise Strongly disagree 1-5 Strongly agree*
- *Sarcasm Strongly disagree 1-5 Strongly agree*

The emotion sets used in the experiment are joy, praise and sarcasm. Joy represents a general positive emotion. Praise and sarcasm are opposites. We set these two options to observe how participants choose and distinguish between positive (praise) and negative (sarcasm) emotions that may be represented in an emoji.

**Results**

In the result analysis of the first part and the second part are overall analysis of the sample data. A specific analysis of the top emoji by age group will be performed in Analysis combined with age.

**The First Part Analysis**

Figure 1

Emoji advised to use ranking

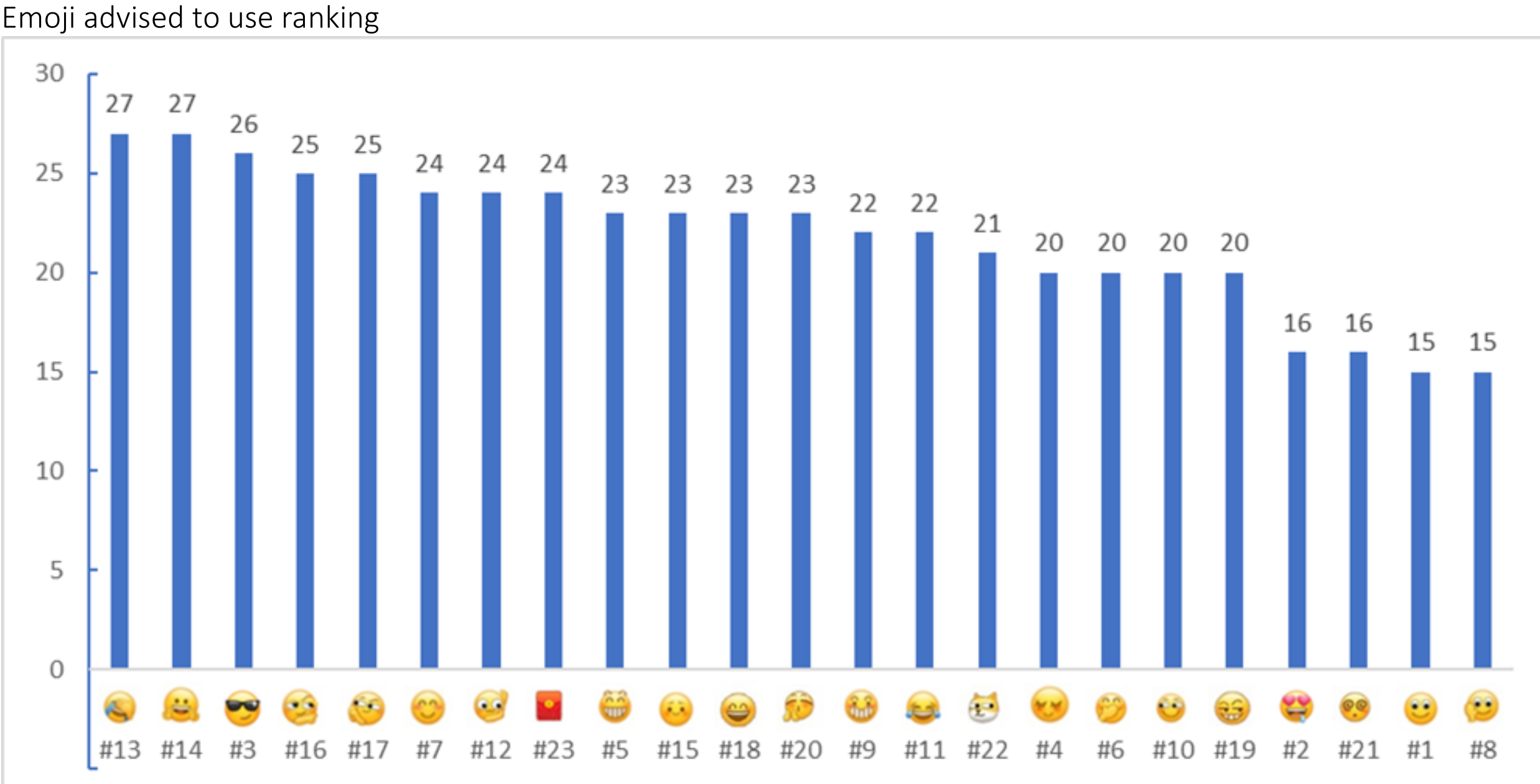

*Note.* The Y-axis is the number of people who chose option a) It can be used with confidence in most cases, without causing misunderstanding.

Among the sample of 29 participants, the top five emojis considered to be acceptable to use without misunderstanding in most cases were [Facepalm], [Yeah!](27). [CoolGuy](26); [Smart],[Respect](25).

Figure 2

Emoji not advised to use ranking

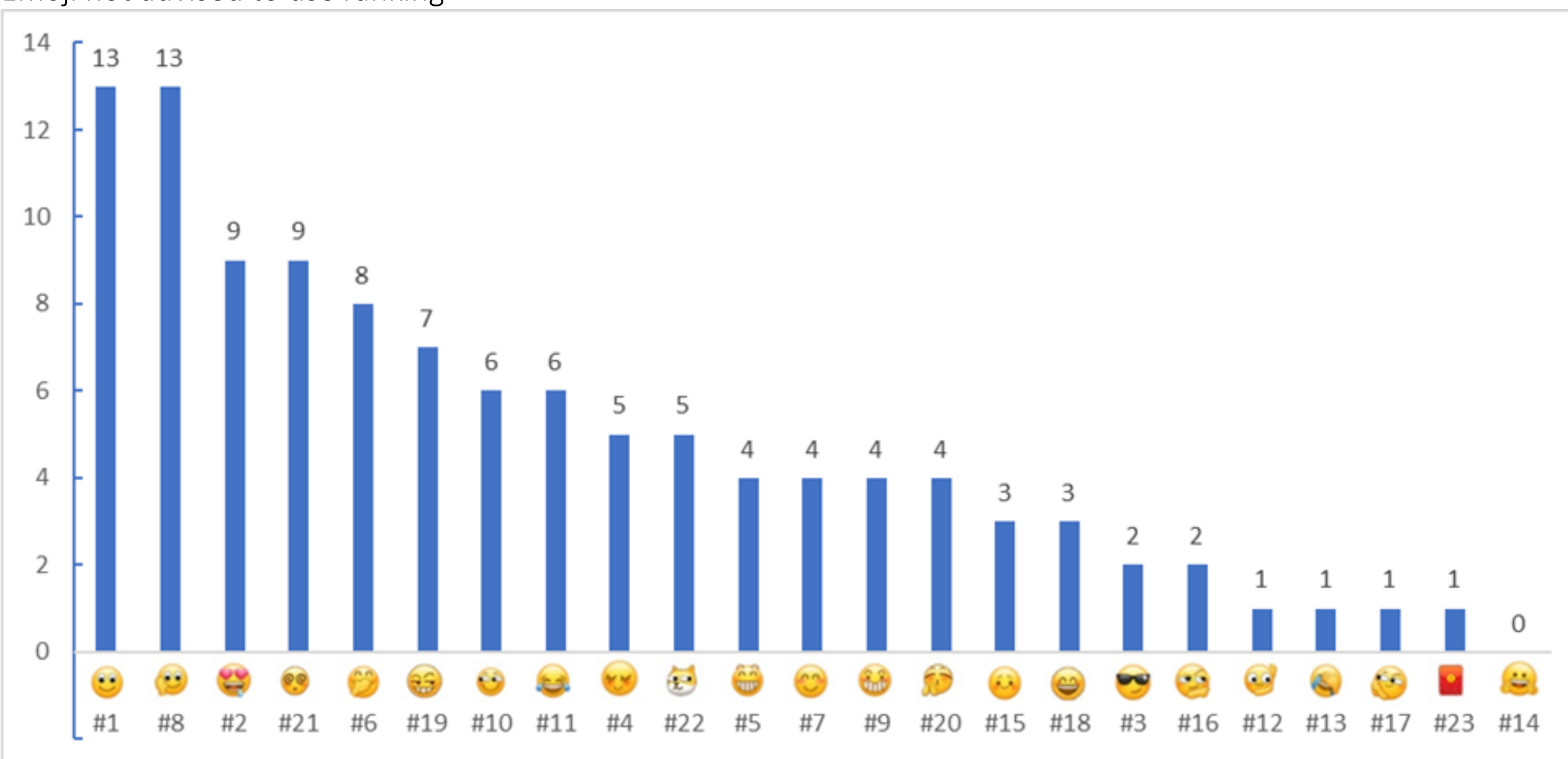

*Note.* The Y-axis is the number of people who chose option b) Do not use the emoji unless the user is very sure of his or her intention in using it.

Among the non-recommended emoji in the sample of 29 participants, [Smile] and [Bye](13) tied for first place, four choices higher than the next best emoji, [Drool](9). This is consistent with previous research and common perception. Ignoring filler [Dizzy], the last three of the top five are [Drool](9); [Chuckle](8); [Sly](7).

**The Second Part Analysis**

Figure 3

Joy mean value for emojis

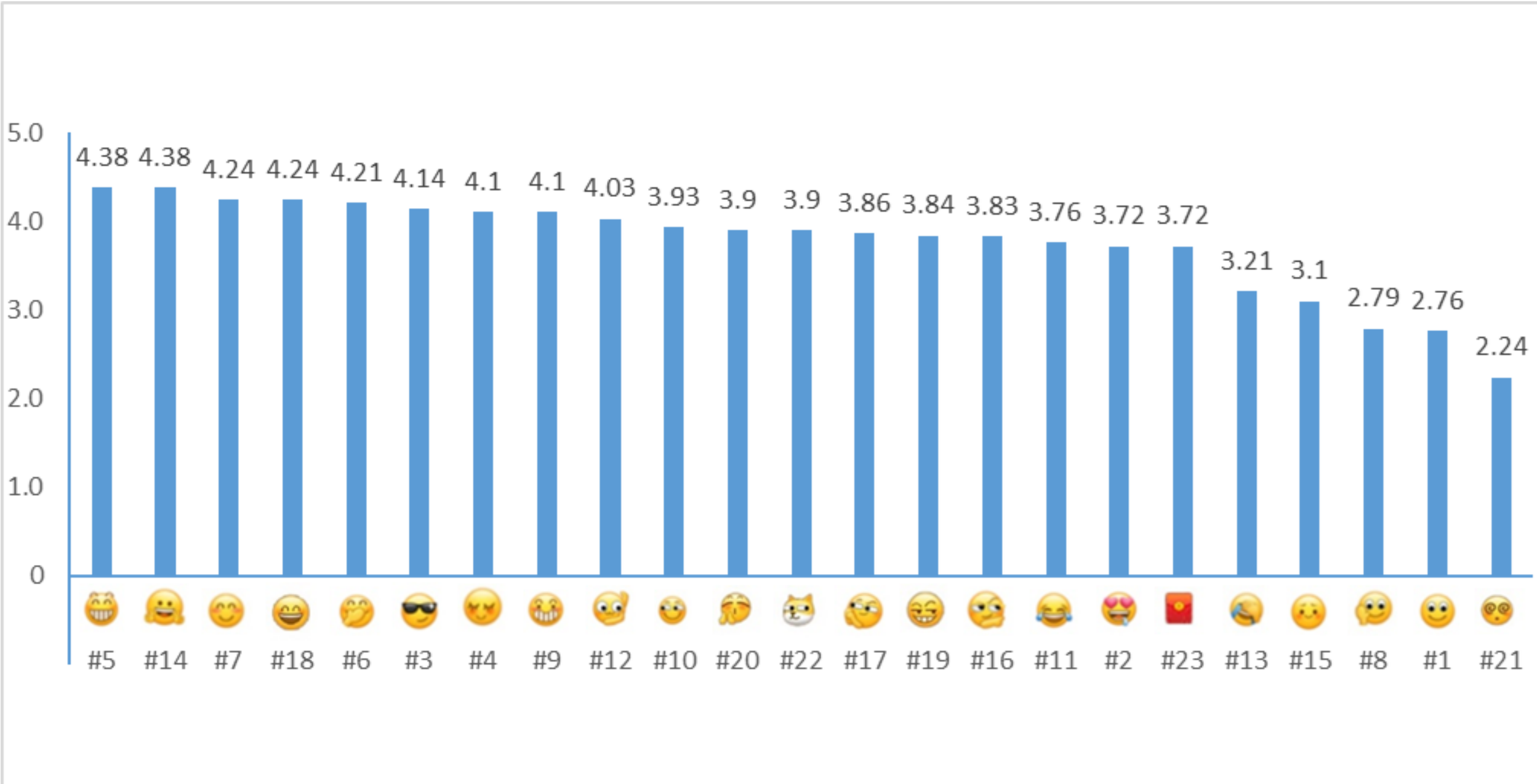

In the sample of 29 participants, the top five emoji considered to express Joy were, [Grin],[Yeah! ][Joyful] [Happy] [Chuckle].

Figure 4

Praise mean value for emojis

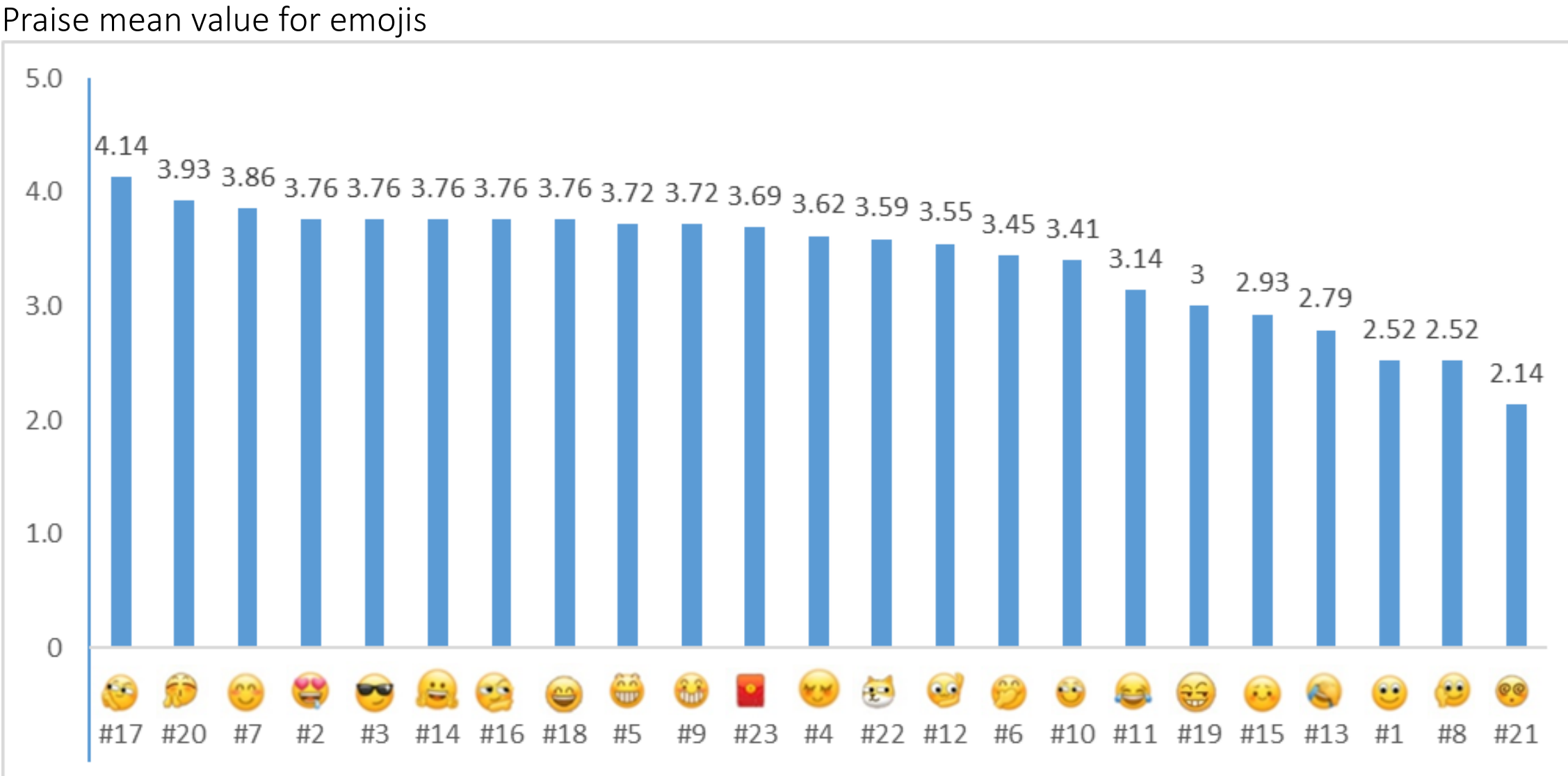

In the sample of 29 participants, the top 5 emoji considered to express praise were, [Respect],[Clap],[Joyful],[Drool], and [CoolGuy].

Figure 5
Sarcasm mean value for emojis

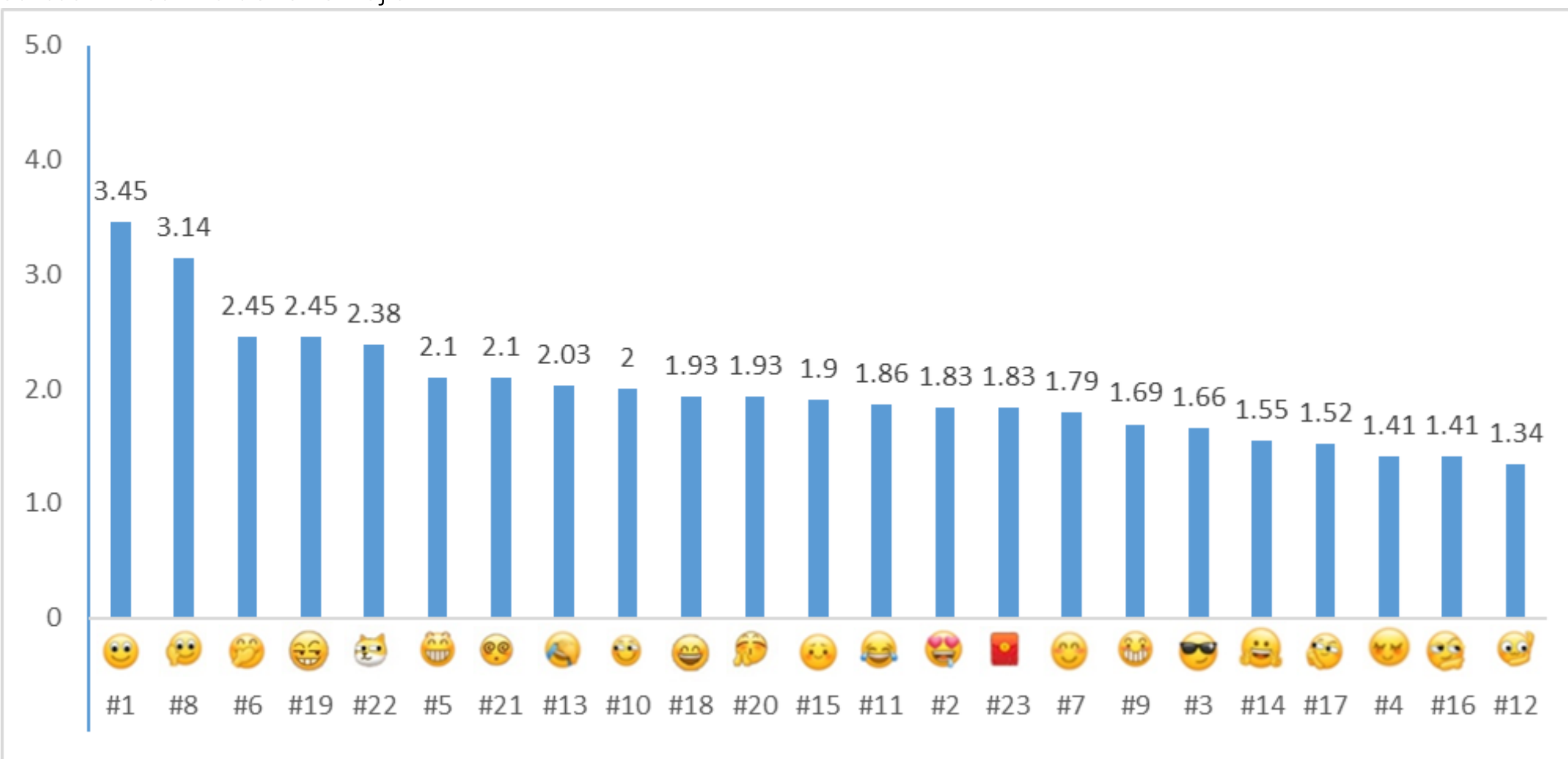

In the sample of 29 participants, the top 5 emoji considered to express Sarcasm were, [Smile],[Bye],[Chuckle],[Sly], and [Doge]. [Doge] is a filler, but in this case, it can be seen that this expression can indeed express sarcasm in many people's eyes. Moreover, [Doge] is 0.28 higher than the emoji score after it, which is a high score difference for this set of emojis. It can be noticed that the top two emojis [Smile] and [Bye] reach more than 3 which are much higher scores than the other emojis. This is consistent with the results in the first part, people believe that [Smile] and [Bye] represent the meaning of sarcasm, so it is not recommended for others to use these two expressions to avoid accidentally making the other side understand the meaning of sarcasm when people do not want to be sarcastic others.

**Analysis Combined with Age**

In this study, age groups were divided into two groups, younger people aged 18-25 years and older people aged over 40 years.

First, we measured the self-assessment of emoji proficiency in different age groups on a scale of 1-10. The mean score was 8.43(SD=1.65) for younger people and 9.38(SD=0.7) for older people. It can be seen that older people think they are familiar with emoji, they are confident in using emoji, and the standard deviation is smaller, indicating that their views are more similar. Among younger people, they also think they are familiar with emoji, but they are not as confident as older people, and the standard deviation is larger, which indicates that some people are more confident, while some are not confident and think they are not familiar with emoji.

Table 3
Age group ratings for [Smile] and [Bye]

| Age group | Emoji | Emotion | avg | SD |
|---|---|---|---|---|
| Younger people (N=21) | Smile | joy | 2.05 | 1.36 |
| | | praise | 1.9 | 1.38 |
| | | sarcasm | 4.29 | 0.93 |
| | Bye | joy | 2.14 | 1.36 |
| | | praise | 1.86 | 1.25 |
| | | sarcasm | 3.86 | 1.04 |
| Older people (N=8) | Smile | joy | 4.63 | 0.7 |
| | | praise | 4.13 | 1.05 |
| | | sarcasm | 1.25 | 0.66 |
| | Bye | joy | 4.75 | 0.66 |
| | | praise | 4.5 | 0.87 |
| | | sarcasm | 1.25 | 0.66 |

Here are the two least recommended emojis [Smile] and [Bye]. As can be seen from the table (Table 3), younger people and older people have almost completely opposite understandings of [Smile] and [Bye]. Younger people rated the positive emotion (joy, praise) of the two emojis lower and the sarcasm higher. Older people scored higher on positive emotions (joy, praise) and lower on sarcasm. In general, older people's standard deviation is smaller, and older people's evaluation score difference for these emoji is smaller.

It can be concluded that older people are confident in their emoji usage, but their interpretation of some emojis, such as [Smile] and [Bye], is opposite to younger people. In this case, when they are communicating, emoji can be easily misunderstood.

Table 4
Age group ratings for [Facepalm] and [Yeah!]

| Age group | Emoji | Emotion | avg | SD |
|---|---|---|---|---|
| Younger people (N=21) | Facepalm | joy | 3 | 1.23 |
| | | praise | 2.43 | 1.29 |
| | | sarcasm | 2.33 | 1.12 |
| | Yeah! | joy | 4.24 | 0.68 |
| | | praise | 3.47 | 1.33 |
| | | sarcasm | 1.48 | 0.73 |
| Older people (N=8) | Facepalm | joy | 3.75 | 1.71 |
| | | praise | 3.75 | 1.71 |
| | | sarcasm | 1.25 | 0.66 |
| | Yeah! | joy | 4.75 | 0.66 |
| | | praise | 4.5 | 0.87 |
| | | sarcasm | 1.75 | 1.39 |

Here are two of the most recommended emojis to use [Facepalm] and [Yeah!]. As can be seen from the table (Table 4), [Yeah!] in younger and older people, the understanding is similar, positive emotion (joy, praise) is rated higher, but sarcasm is rated lower. [Facepalm] belongs to positive emotion for older people, positive emotion (joy, praise) has a high score, and sarcasm has a low score. However, for younger people, [Facepalm] belongs to a neutral position, joy, praise, and sarcasm are not particularly high, and the understanding of this expression may be complicated and largely depends on the context.

## Conclusion

It can be seen from this study that people's emoji usage habits are sometimes consistent with their intuition, but also often not. Older people feel confident in using emoji such as [Smile], but if they are sending this emoji to younger people, younger people are more likely to come up with the opposite meaning, leading to misunderstandings.

For some emoji, such as [Yeah!] understanding is similar in different age groups and using such expressions can reduce misunderstandings.

Understanding the emoji usage habits of online culture and different people is useful to reduce emoji misunderstandings. People can be surer that the emojis sent will not cause misunderstanding, and sending emojis to friends that they can understand more can reduce misunderstandings, thus improving the communication effect and efficiency.

For the emoji recommendation system, if the user is willing to provide some information about the message receiver, the recommendation system can more accurately recommend emojis that the receiver can understand easier. Or the system can recommend emoji that are not easy to create ambiguity.

Future studies could increase the sample size and obtain enough samples at each age to analyze people's emoji usage habits. People's usage habits and Internet culture are evolving, and it is meaningful to continuously track the usage habits and evolution process of emoji through long-term research, which can enable people to have a deeper and systematic understanding of emoji and Internet culture.